\documentclass[a4paper,12pt]{article}
\usepackage{amssymb}
\usepackage{graphicx}
\usepackage{amsmath}
\usepackage{amsfonts}

\begin{document}

\title{Dynamical bag in a chiral quark model}
\author{Duojie Jia$^{\thanks{%
E-mail: jiadj@nwnu.edu.cn; }\ \dagger }$,LianChun Yu$^{\ddagger }$, Rui-Bin
Wan$^{\ast }$ \\
%EndAName
$^{\ast }$Institute of Theoretical Physics, College of Physics and \\
Electronic Engineering, Northwest Normal University,\ \\
Lanzhou 730070, China\\
$^{\ddag }$Institute of Theoretical Physics, College of Physical Science\\
and Technology, Lanzhoul University, Lanzhou 730000, China}
\maketitle

\begin{abstract}
A type of bag function is proposed to make the MIT bag surface of baryon
dynamical. It is illustrated through renormalization of the quark field that
the softening of chiral bag gives rise to a model of chiral quark with
effectively-generated mass of quark, in which confined quark moves in the
background of nonlinear pion. A prediction of bag constant $B$ $\simeq
2f_{\pi }^{2}m_{\pi }^{2}\allowbreak $ is made. With two free parameters,
the self-coupling $e$ of pion and the confining scale $a$, the computed
mass, the charge root-mean-square radius and magnetic moment of the proton
are in good agreement with the experimental values.

PACS number(s): 12.39.Ba,12.39.-x,14.20.-c

\end{abstract}

\section{Introduction}

In principle, the fundamental theory of strong interaction, the Quantum
Chromodynamics (QCD), provides the whole information of the hadrons from the
underlying degrees of freedom, the quark and gluon. The complexity of QCD
vacuum, however, makes the direct calculation via QCD, e.g., the simulation
via QCD on lattice, quite nontrivial for the light sector of quark\cite{P.
Hagler}(below $\sim 1GeV$).

On the other hand, the effective models of hadrons, such as the Skyrme model%
\cite{Skyrme,ANW}, the MIT bag models \cite{MIT0,DeGrand75}, the cloudy bag
model\cite{CBM}, the quark models in the nonrelativistic form(NQM)\cite%
{IsgurKarl} and the relativistic form \cite{PCQM,Faessler}, etc., also
enable us to explore the strong interaction at low-energy. While the NQM is
successful, the MIT bag model provides a conceptually simple picture of
hadrons. Though it incorporates both confinement and asymptotic freedom of
QCD, the MIT bag model, in its original form\cite{MIT0}, did not possess
chiral symmetry, another main feature of QCD at low energy. In the hybrid
chiral bag(HCB) model, Chodos and Thorn\cite{Chodos75} and Brown and Rho
\cite{BrownRho} restored chiral symmetry by introducing pions into the bag
Lagrangian. Thus a baryon can be thought to be three-quarks inside a bag,
surrounded by a cloud of pions, which makes the nucleon-nucleon\ interaction
description\cite{BrownRho} via the exchange of pions accessible.

In the nonlinear-pion treatment the HCB model\cite{Chodos75,JAC87} (see\cite%
{Hosaka96} for a review) the confined quarks inside bag interact with the
pion through the chirally-invariant coupling at the bag surface by
interpolating the Skyrme model in the limit of the small-bag(the bag radius $%
R\rightarrow 0$) and the MIT bag model in the large-bag limits($R\rightarrow
\infty $). To do so, a bag function $\theta _{V}$(unit in the bag volume $V$
and zero outside) and its delta function $\delta _{S}$($=\infty $ on the
surface $S$ of $V$ and zero elsewhere) were used to imitate the permanent
confinement of quarks in hadrons and ensure the continuity of the
axial-vector current. The HCB Lagrangian reads\cite{Hosaka96}
\begin{equation}
\mathcal{L}^{HCB}=\left( \bar{q}i/\kern-0.57em \partial q-B\right) \theta _{V}-\frac{1%
}{2}\bar{q}U_{5}q\delta _{V}+\mathcal{L}^{\pi }\theta _{\bar{V}},
\label{HCB}
\end{equation}%
where $\mathcal{L}^{\pi }$ is the Lagrangian of pion $\mathbf{\pi }$, $%
U_{5}=\exp (i\boldsymbol{\tau }\cdot \mathbf{\pi }\gamma _{5}/f_{\pi })$, $%
\theta _{\bar{V}}=1-\theta _{V}$. In the case that $\mathcal{L}^{\pi }$ is
the Skyrme Lagrangian\cite{Hosaka96} the model (\ref{HCB}) describes a quark
bag as a defect in the Skyrme field, and realizes the chiral symmetry in the
Wigner mode inside the bag and in the Nambu-Goldstone mode outside, with the
baryon charge ($B$) conserved topologically. Due to the sharp surface of the
bag which allows the creation of virtual pions with arbitrarily high
momenta, however, the self-energy of nucleon remains to be infinite and the
origin of the bag to be unknown in the bag models including the HCB model (%
\ref{HCB}).

One way to avoid self-energy divergence of nucleon is to include the
exterior fermions\cite{Vepstas} with the finite self-energy of nucleon
arising from Casimir effect. Another approach\cite{PCQM,Faessler} is to
replace the unphysical bag boundary by a finite surface thickness of the
quark core within the framework of the relativistic quark model with static
potential. Also, an attempt \cite{Nogami} was made to regularize the
self-energy by introducing a soft(fuzzy) bag surface.

In this paper, we propose a type of soft bag function formed by nonlinear
pion to make the bag surface of baryon dynamical. It is illustrated that the
softening of chiral bag gives rise to a chiral quark model with an
effectively generated mass of quark, in which confined valence quark
interacts with a chiral cloudy of the pion. The finite mass of nucleon
including pion dressing is obtained through renormalization of the quark
field in terms of pion, and a prediction $B$ $\simeq 2f_{\pi }^{2}m_{\pi
}^{2}$ for the bag constant is made$\allowbreak $. Using two free parameters
of the model, the self coupling $e$ of self-pion and the confining scale $a$%
, some properties(the mass, the charge root-mean-square radius and magnetic
moment) of the proton are computed, in good agreement with the experimental
values. The bag radius and the size of quark core are estimated to be about $%
0.91fm$ and $0.3fm$, respectively. This is done by using a trial profile for
the hedgehog Skyrmion\cite{JiaWang} in the process of solving model.

\section{The quarks in a cloudy pion}

To incorporate the surface dynamics into the HCB model in which the bag
functions($\theta _{V},\delta _{S}$) are singular, we propose the following
continuous bag functions constructed in terms of the nonlinear chiral field $%
U(r)$ of pion, to replace the singular bag functions in the HCB model,
\begin{equation}
\begin{array}{c}
\theta _{V}\rightarrow \theta _{U}=\frac{1}{8}tr[2-(U+U^{\dagger })], \\
\delta _{S}\rightarrow \delta _{U}=-\frac{1}{8}\frac{d}{d\hat{n}}%
tr[2-(U+U^{\dagger })], \\
\theta _{\bar{V}}\rightarrow \theta _{\bar{U}}=\frac{1}{8}tr[2+(U+U^{\dagger
})]%
\end{array}
\label{SDel}
\end{equation}%
where $U=\exp (i\boldsymbol{\tau }\cdot \boldsymbol{\pi }/f_{\pi })$ and $d/d%
\hat{n}$ stands for the gradual derivatives along the direction normal to
the bag surface($S$). Here, $U$ is the nonlinear representation of the
Goldstone bosons($\pi $ mesons) under the constraint $U^{\dag }U=1$. To see
how $\theta _{U}$ and $\delta _{U}$ behave as the soft bag functions, we
utilize as a first step a simple Skyrmion profile in the Skyrme model for
the chiral angle
\begin{equation}
F(r)\simeq 4\arctan [\exp (-c_{0}r)],  \label{FSK}
\end{equation}%
with $r$ the radial coordinate measured in the unit of $(ef_{\pi })^{-1}$
(the parameters in the Skyrme Lagrangian, see (\ref{SK}) below). It is shown
numerically\cite{Sutcliffe,JiaWang} that a Skyrmion in finite region can be
approximated by (\ref{FSK}) $c_{0}\simeq 1.0$. With the hedgehog ansatz for $%
U=\exp [iF(r)\hat{r}\cdot \boldsymbol{\tau }]\in SU(2)$ in (\ref{SDel}), one
finds%
\begin{equation}
\begin{array}{c}
\theta _{U}=\frac{1}{2}\left( 1-\cos (F)\right) =\sin ^{2}(F/2)\rightarrow
\sec h^{2}[c_{0}r], \\
\theta _{\bar{U}}=\frac{1}{2}\left( 1+\cos (F)\right) =\cos
^{2}(F/2)\rightarrow \tanh ^{2}[c_{0}r], \\
\delta _{U}=-\frac{1}{2}\sin (F)\frac{dF}{dr}\rightarrow 2c_{0}\sec
h^{2}(c_{0}r)\tanh (c_{0}r).%
\end{array}
\label{SD2}
\end{equation}%
Here, the symbol arrow($\rightarrow $) refers the further reduction upon
using (\ref{FSK}), which are plotted explicitly in \textrm{FIG.1}.%

\begin{figure}[ptb]\begin{center}
\includegraphics[width=0.8\textwidth]{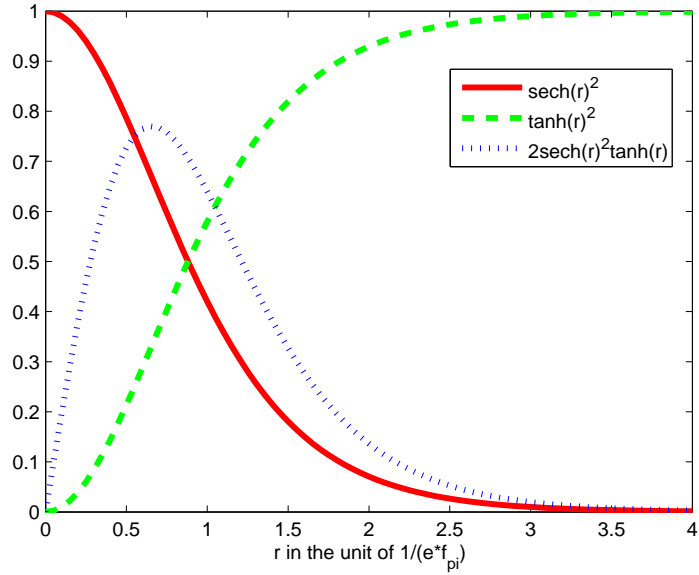}
 \caption{The three functions of soft bag.
}\label{Bagfun}
\end{center}\end{figure}

Imposing the (chiral) boundary condition $F(0)=\pi ,F(\infty )=0$,
corresponding to the physical vacua $U=\pm 1$, respectively, one can check
from (\ref{SDel}) alone that the relations $\theta _{\bar{U}}=1-\theta _{U}$
and $\int_{0}^{\infty }dr\delta _{U}=1$ hold, just as $\theta _{V}$ and $%
\delta _{S}$ do in the HCB model. We note that these relations hold simply
as a pure result of the boundary condition on $F$, with nothing to do with
the explicit profile (\ref{FSK}) for the Skyrmion. Given the success of the
Skyrme model in description of the light baryons below $1$ GeV, it is
natural to extend the solitonic profile (\ref{SD2}) into a bag-like
functions constructed by the dynamical degree of freedom, $U$.

We consider the following reformulation of the HCB model (\ref{HCB}) with
the replacement (\ref{SDel}) used and a scalar-like confining potential $S(x)
$ added,
\begin{equation}
\mathcal{L}^{DCB}=[\bar{q}(i\gamma ^{\mu }\partial _{\mu
}-S(x)U_{5})q-B]\theta _{U}-\frac{1}{2}\bar{q}U_{5}q\delta _{U}+\mathcal{L}%
^{SK}(U)\theta _{\bar{U}},  \label{MCB}
\end{equation}%
in which $q\,$is the current quark of light flavor, $\mathcal{L}^{SK}(U)$
describes the chiral dynamics of the pion in terms of nonlinear chiral field
$U$ (or equivalently, of $F$). Upon using (\ref{SDel}) in (\ref{MCB}), one
finds
\begin{equation}
\begin{array}{c}
\mathcal{L}^{DCB}=\frac{1}{2}[\bar{q}(i\gamma ^{\mu }\partial _{\mu
}-SU_{5})q]-\frac{1}{8}tr(U+U^{\dag })[\bar{q}(i\gamma ^{\mu }\partial _{\mu
}-SU_{5})q] \\
-\frac{1}{16}\bar{q}U_{5}q\left( n^{\mu }tr(\partial _{\mu }U+\partial _{\mu
}U^{\dag })\right) +\mathcal{L}^{\pi }%
\end{array}
\label{DCB}
\end{equation}%
in which $n^{\mu }=(0,\hat{r})$. Assuming the pion Lagrangian $\mathcal{L}%
^{SK}(U)$ to be, for instance, the twice of Skyrme model $2\mathcal{L}%
^{Skyrme}$(this corresponds to redefinition of the model parameters) one has
explicitly for the pion dynamics in (\ref{DCB}),%
\begin{equation}
\begin{array}{c}
\mathcal{L}^{\pi }=\frac{f{}_{\pi }^{2}}{4}tr(\partial _{\mu }U\partial
^{\mu }U^{\dag })+\frac{1}{8}Btr[U+U^{\dagger }-2] \\
+\frac{f{}_{\pi }^{2}}{16}tr[U+U^{\dagger }]tr[\partial _{\mu }U^{\dag
}\partial ^{\mu }U]+\frac{1}{16e^{2}}tr[(\partial _{\mu }U^{\dag }\partial
_{\nu }U)^{2}-(\partial _{\mu }U\partial ^{\mu }U^{\dag })^{2}] \\
+\frac{1}{64e^{2}}tr[U+U^{\dagger }]tr[(\partial _{\mu }U^{\dag }\partial
_{\nu }U)^{2}-(\partial ^{\mu }U^{\dag }\partial _{\mu }U)^{2}]].%
\end{array}
\label{piondyn}
\end{equation}%
\newline
Here, the Skyrme Lagrangian is\cite{Skyrme}
\begin{equation}
\mathcal{L}^{SK}=\frac{f{}_{\pi }^{2}}{4}tr(\partial _{\mu }U\partial ^{\mu
}U^{\dag })+\frac{1}{32e^{2}}tr[\partial _{\mu }UU^{\dag },\partial _{\nu
}UU^{\dag }]^{2},  \label{SK}
\end{equation}%
with $f_{\pi }$ the pion decay constant, and $e$ the self-coupling of the
pion.

The model (\ref{DCB}) describes an interacted theory of valence quark and
pion, which breaks chiral symmetry through the coupling the quark bilinear
operators to the terms $tr(U+U^{\dag })$ and $tr(\partial _{\mu }U+\partial
_{\mu }U^{\dag })$. We note here that the pion dynamics (\ref{piondyn}) is
closed upon the replacement (\ref{SDel}) in the sense that it simply picks
up some of the higher terms (up to the order-six terms of the momentum
expansion) from the chiral perturbative theory(ChPT)\cite{ChPT} without
extra parameters added. By comparing (\ref{piondyn}) with the ChPT, one
finds $B\simeq 2f_{\pi }^{2}m_{\pi }^{2}$, which results in the bag constant
$B\simeq (134MeV)^{4}\allowbreak $(with the pion mass $m_{_{\pi }}\simeq
137MeV$), in agreement with $B^{1/4}=146MeV\allowbreak $ in the MIT bag
model \cite{DeGrand75}.

The difference of the bag constant estimated here with that of the MIT bag
model and the value $B^{1/4}\simeq 150MeV$ in chiral bag model\cite{Hosaka96}
may be due to the pion self-coupling near the bag region. In fact, the
estimate $B\simeq 2f_{\pi }^{2}m_{\pi }^{2}$ is made through the chiral(weak-%
$\pi $) expansion of $U$: $U\approx 1+i\pi /f_{\pi }-(\pi /f_{\pi })^{2}/2$.
In fact, the chiral(weak-$\pi $) expansion
\begin{equation*}
\theta _{U}\approx (\pi /2f_{\pi })^{2}-(\pi /2f_{\pi })^{4}/3
\end{equation*}%
does not apply near the bag region(where $\theta _{U}\sim 1$, see \textrm{%
FIG.1}), as it ignores the strong self-interaction of the nonlinear pion
there.

\section{The renormalization of quark by pion}

We see that the model (\ref{DCB}), as it stands, does not assume the
standard form of relativistic field theory. To see the role of pion clearly,
it is very useful to reformulate (\ref{DCB}) into the form of relativistic
field theory by changing the quark field $q$ into a new variable of
effective quark, that is normalized.

We note first that the MIT wavefunction $q_{MIT}$, when normalized as $\int
q_{MIT}^{\dag }q_{MIT}d^{3}x=1$, can be written in whole space as
\begin{equation}
q_{MIT}=\psi _{\kappa m}(x)\theta (R-r)^{1/2}  \label{MIT}
\end{equation}%
where $\psi _{\kappa m}$ is the solution to the MIT bag model. The form (\ref%
{MIT}) is also valid outside the bag($r>R$) since $\theta (R-r)$
vanishes there. The wavefunction (\ref{MIT}) can also be obtained by
solving an equivalent problem\cite{Mosel} $\mathcal{L}^{MIT}=q[i
/\kern-0.57em {\partial}-m(r)]q-B$ in an infinite vallum potential
$m(r)$, namely, the Dirac equation for the quark having the
effective vanishing mass $m(r)=0$(for $r\leqslant R$) and the
infinite mass ($m(r)=\infty ,$for $r>R$).

The similar picture of the effective mass occurs for our model, if one
redefines a renormalized quark as\cite{Nogami}
\begin{equation}
Q=\theta _{U}^{1/2}q=\left\{ \frac{1}{8}tr[2-(U+U^{\dagger })]\right\}
^{1/2}q  \label{Qq}
\end{equation}%
and requires $Q$ to fulfill the reorganization $\int Q^{\dag }QdV=1$ as $%
q_{MIT}$ do in the MIT bag model. Upon substitution of (\ref{Qq}) into the (%
\ref{MCB}), one finds
\begin{equation}
\begin{array}{c}
\mathcal{L}^{CQ}=\bar{Q}[i\gamma ^{\mu }(\partial _{\mu }-\frac{1}{2}%
\partial _{\mu }\ln \theta _{U})-SU_{5}+\frac{1}{2}U_{5}n^{\mu }\partial
_{\mu }\ln \theta _{U}]Q \\
+\mathcal{L}^{\pi }%
\end{array}
\label{CQ1}
\end{equation}%
where the pion dynamics $\mathcal{L}^{\pi }$ is given by (\ref{piondyn}).

The static Hamilton associated with (\ref{CQ1}) is%
\begin{equation}
\begin{array}{c}
H^{CQ}=\int d^{3}xQ^{\dag }\left\{ -i\mathbf{\alpha }\cdot (\mathbf{\nabla }-%
\hat{r}\cot (F/2)F_{r}/2)\right. \\
\left. +\gamma ^{0}(M(r)+S(r))\cos F\right\} Q+B\int d^{3}x\sin
^{2}(F/2)+H^{\pi },%
\end{array}
\label{Hmodel2}
\end{equation}%
with
\begin{equation}
M(r)\equiv \cot (F/2)(-F_{r}/2),  \label{Mr}
\end{equation}%
and $F_{r}=dF/dr$. Here, in this work, the Hamilton $H^{\pi }$ for the pion
dynamics is truncated, for simplicity, to be that of the Skyrme model(three
foregoing terms) in (\ref{piondyn})
\begin{equation}
\begin{array}{c}
H^{\pi }=\int d^{3}x\left\{ \frac{f{}_{\pi }^{2}}{4}tr(\partial
^{i}U\partial ^{i}U^{\dag })+\frac{1}{4}Btr[1-U]\right. \\
+\left. \frac{1}{16e_{s}^{2}}tr[(\partial _{i}U^{\dag }\partial
_{i}U)^{2}-(\partial _{i}U^{\dag }\partial _{j}U)^{2}]\right\}%
\end{array}
\label{Hpion}
\end{equation}

Taking the wave function for the (valence) quark $Q$ to be
\begin{equation}
\psi _{Q}=\frac{N}{r}\left(
\begin{array}{c}
G(r) \\
-iH(r)\boldsymbol{\sigma }\cdot \hat{r}%
\end{array}%
\right) \emph{y}_{ljm}(\theta \varphi )\chi _{f}  \label{psiQ}
\end{equation}%
with $N^{2}\equiv 1/[\int dr(G^{2}+H^{2})]$ the normalization factor, $\emph{%
y}_{ljm}$ the\ Pauli spinor and $\chi _{f}$ the flavor wavefunction, one
finds for (\ref{Hmodel2})
\begin{equation*}
\begin{array}{c}
H^{CQ}=N^{2}\int dr\left\{ H\frac{dG}{dr}-G\frac{dH}{dr}+\frac{2\kappa }{r}%
GH+(M+S)\cos F(G^{2}-H^{2})\right\}  \\
+4\pi B\int drr^{2}\sin ^{2}(F/2)+H^{\pi },%
\end{array}%
\end{equation*}%
with $-\kappa $ the eigenvalue of the operator $K=\gamma ^{0}[\mathbf{\Sigma
}\cdot (\mathbf{r}\times \mathbf{p})+1]$ corresponding to the eigen-state $%
\emph{y}_{ljm}$, and it, upon the re-scaling $r=Lz(L=1/ef_{\pi })$, becomes%
\begin{equation}
\begin{array}{c}
H^{CQ}=N^{2}\int dz\left\{ H\frac{dG}{dz}-G\frac{dH}{dz}+\frac{2\kappa }{z}%
GH+(LM+LS)\cos F(G^{2}-H^{2})\right.  \\
+\left. \frac{4\pi L^{4}B}{LN^{2}}z^{2}(1-\cos F)\right\} +E^{\pi },%
\end{array}
\label{HCQ2}
\end{equation}%
with $N^{2}L=1/\int dz[G^{2}+H^{2}]$, and $E^{\pi }$ given by
\begin{equation}
\begin{array}{c}
E^{\pi }=\frac{2\pi f_{\pi }}{e}\int dz\left\{ z^{2}F_{z}^{2}+2\sin
^{2}(F)(1+F_{z}^{2})\right.  \\
\left. +\frac{\sin ^{4}F}{z^{2}}+e^{2}L^{4}Bz^{2}(1-\cos F)\right\}
\end{array}
\label{Ep}
\end{equation}%
where $F_{z}\equiv dF/dz$. The equation of motion for (\ref{HCQ2}) with (\ref%
{Ep}) is
\begin{equation}
\begin{array}{c}
\frac{dG}{dz}+\frac{\kappa }{z}G=[\varepsilon +L(M+S)\cos F]H \\
-\frac{dH}{dz}+\frac{\kappa }{z}H=[\varepsilon -L(M+S)\cos F]G \\
\left( 1+\frac{2\sin ^{2}F}{z^{2}}\right) F_{zz}+\frac{2\sin (2F)}{z}\left(
F_{z}^{2}-1-\frac{\sin ^{2}(F)}{z^{2}}\right)  \\
=e^{2}\left[ L^{4}B-\frac{LN^{2}}{4\pi }(LM+LS)\frac{(G^{2}-F^{2})}{z^{2}}%
\right] \sin F%
\end{array}
\label{EQM}
\end{equation}%
where $\varepsilon =E_{q}L$ is the eigen-energy of $Q$, and
\begin{eqnarray*}
LM(z) &=&\cot (F/2)(-F_{z}/2). \\
LS(z) &=&L(r/a^{2})=L_{a}^{2}z,
\end{eqnarray*}%
with $L_{a}=L/a=1/(ef_{\pi }a)$. Here we have taken the scalar potential $S$
to be of the linear form $S=r/a^{2}$, with $a$ the confining scale.

It follows from the asymptotic form of (\ref{EQM}) that
\begin{equation}
\begin{array}{c}
G(z\rightarrow \infty )\sim e^{-L_{a}^{2}z^{2}/2}\sim -H(z\rightarrow \infty
), \\
F(z\rightarrow \infty )\sim e^{-2\sqrt{B}L^{2}z}/z,%
\end{array}
\label{Asmi}
\end{equation}%
and
\begin{equation}
\begin{array}{c}
G(z\approx 0)\approx \sqrt{z}Y_{\kappa +1/2}(\varepsilon z)\sim \sqrt{\frac{%
2\varepsilon }{\pi }}z, \\
H(z\approx 0)\sim -zJ_{\kappa +3/2}(\varepsilon z)\sim -\sqrt{\frac{%
2\varepsilon }{\pi }}z^{3/2}, \\
F(z\approx 0)\approx \pi -Az.%
\end{array}
\label{Asm0}
\end{equation}

One sees from (\ref{Hmodel2}) that an effective($r$-dependent) mass appears,
as a result of the pion dressing in the form of renormalization (\ref{Qq}).
This is similar in the spirit to the equivalent description of the MIT bag
where quark has a $r$-dependent effective mass $m(r)$($\simeq 0$ inside bag
and $\simeq +\infty $ outside), or equivalently, moves in a vallum-like
scalar potential\cite{Mosel}. This is in consistent with the mechanism of
mass generation\cite{NJL} of quark through chiral symmetry breaking of QCD
at low energy. The effective mass (\ref{Mr})\ is plotted in \textrm{FIG.2}%
(a) against $z=ef_{\pi }r$ for the improved configuration (\ref{Y0}) given
below in section 4.

\section{The proton as a 3-quark state}

According to the n\"{a}ive quark model, a nucleon is a three-quark state of
light quark u, d and s, constructed based on $SU(6)$ flavor symmetry. This
picture is valid approximately when the strange(s quark) component in the
nucleon ignored, as is assumed commonly. We check our model (\ref{MCB}) in
the $N_{f}=2$ flavor case which is relevant to the low-lying state of
nucleons. We first solve the coupled equations (\ref{EQM}) for the
quark-pion system and then take the nucleon(proton, here) state as a
three-quark state with the $SU(6)_{F}$ symmetry of spin-flavor wavefunction.
At last, some of static properties of proton are computed.

To solve the coupled equations (\ref{EQM}), the third of which is nonlinear,
we firstly choose the following profile for the Skyrmion\cite{JiaWang} as a
trial function solution to the third equation in (\ref{EQM})%
\begin{equation}
\begin{array}{c}
F(x)=4w\arctan [e^{-cz}]+\pi (1-w) \\
\times \left[ 1-\left( \frac{\sinh ^{2}(dz)}{a_{l}^{2}+\sinh ^{2}(dz)}%
\right) ^{-1/2}\right]%
\end{array}
\label{Y0}
\end{equation}%
and then solve the equation (\ref{EQM}) self-consistently. $F$ is obtained
by averaging $v=(LM+LS)(G^{2}-F^{2})/(4\pi z^{2})$ over the interval $%
[0,x_{N}]$ and applying the optimizing (Neilder-Mead) algorithm used in \cite%
{JiaWang}. For $x_{N}=a/L=7.8011$, the optimal result is $\langle v\rangle
=0.0097$, and that for (\ref{Y0}) is
\begin{equation}
(a_{l},c,d,w)=(0.620,0.904,1.283,1.174).  \label{paraF}
\end{equation}%
The model parameters are set to be%
\begin{equation}
\begin{array}{c}
e=2.80,a=1/0.6676GeV^{-1}, \\
f_{\pi }=93MeV,B^{1/4}=0.130GeV.%
\end{array}
\label{para}
\end{equation}%
only two ($e,a$) of which are free since $B$ is fixed by the proton mass $%
E_{p}=938.27MeV$(input). The boundary condition are fixed by (\ref{Asmi})
and (\ref{Asm0}). We plotted the resulting solution($k=-1$) for the quark
wavefunction ($G,H$) in \textrm{FIG.2}, corresponding to the energy
eigenvalues($E_{q}=\epsilon ef_{\pi }$)
\begin{equation}
\begin{array}{c}
\epsilon =0.9315, \\
E_{q}=242.557MeV,%
\end{array}
\label{BC}
\end{equation}%
and the chiral angle($F$) as well as $v$ in \textrm{FIG.3}. The
corresponding bag functions and the effective mass of quark are shown in
\textrm{FIG.4}.

The bag radius is calculated using the definition $R=F^{-1}(\pi /2)$, and
listed in the \textrm{Table.I} in which the size of the quark core $R_{a}/R=a
$ is fixed by (\ref{Asmi}),$\,$%
\begin{eqnarray*}
R &=&0.9092fm, \\
a &=&\allowbreak 0.2955fm,\,
\end{eqnarray*}%

\begin{figure}[ptb]\begin{center}
\includegraphics[width=0.8\textwidth]{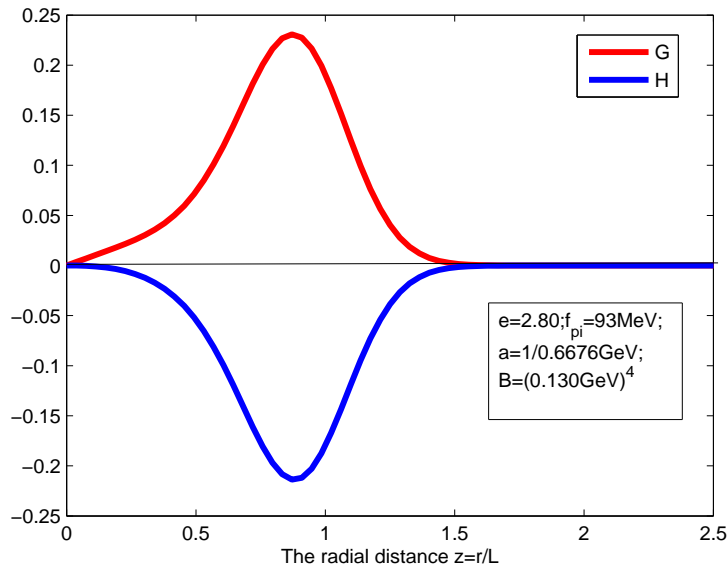}%
\caption{The three functions of soft bag.}
\end{center}\end{figure}%

\begin{figure}[ptb]\begin{center}
\includegraphics[width=0.8\textwidth]{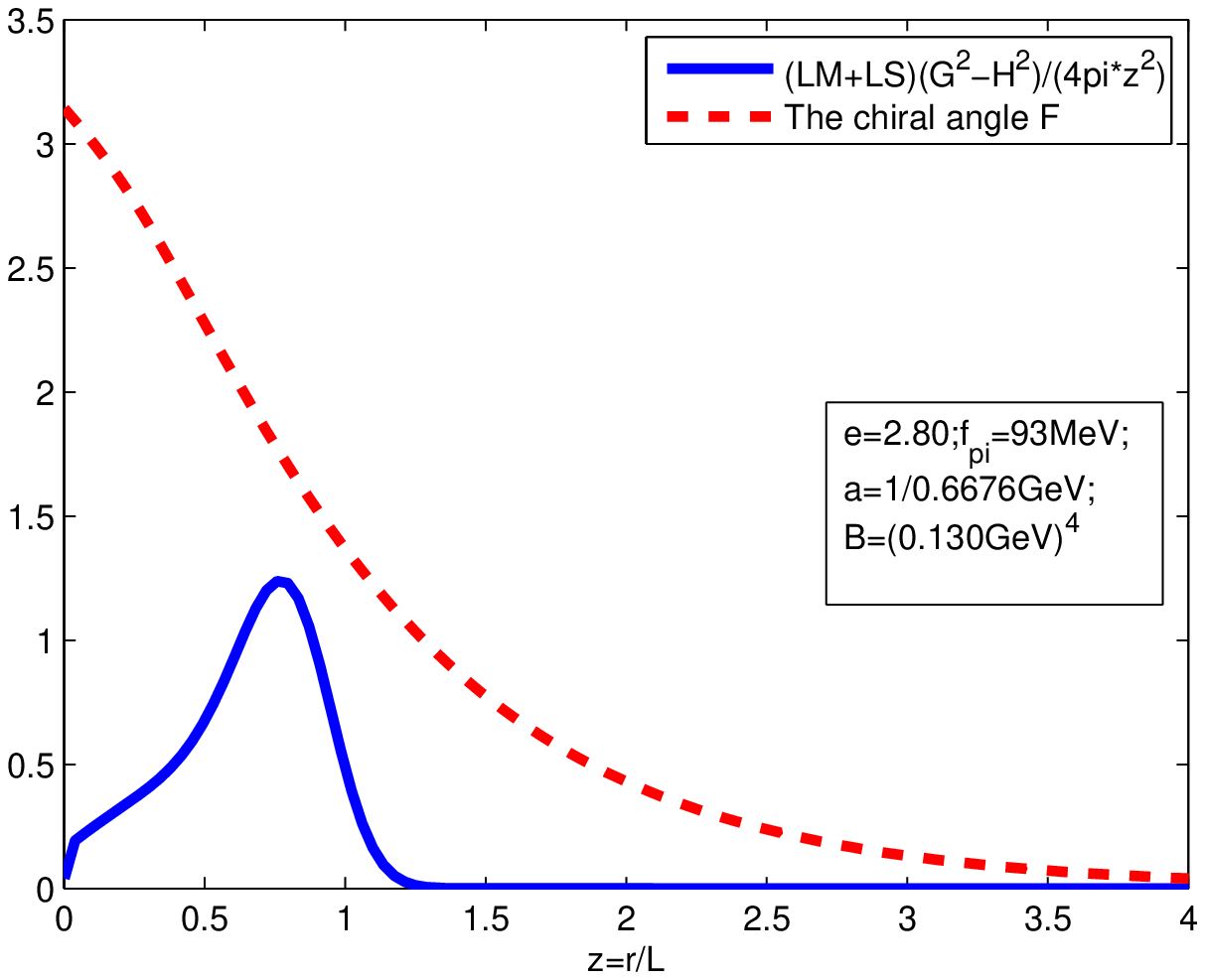}%
\caption{The three functions of soft bag.}
\end{center}\end{figure}%

\begin{figure}[ptb]\begin{center}
\includegraphics[width=0.8\textwidth]{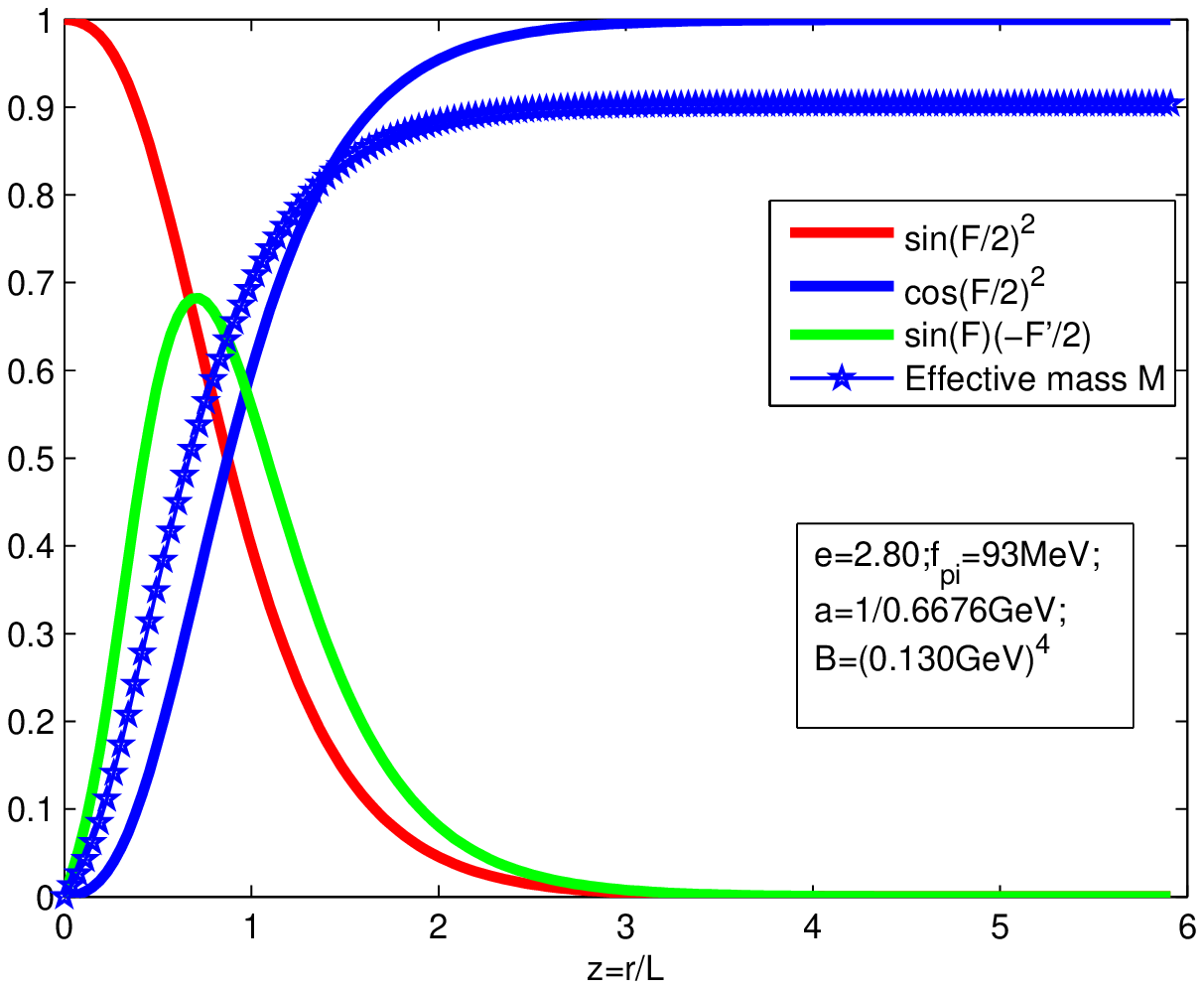}%
\caption{The three functions of soft bag.}
\end{center}\end{figure}%

Following the n\"{a}ive quark model, we write the nucleon state for the
proton to be that of three valence quarks
\begin{equation}
|p\uparrow \rangle =\frac{1}{\sqrt{3}}[|u\uparrow u\downarrow d\uparrow
\rangle _{S}-2|u\uparrow u\uparrow d\downarrow \rangle _{S}],  \label{p}
\end{equation}%
in which the suffix $S$ (here) stands for the symmetrization of the quark
indices($1,2,3$), namely, the position in the ket representation $%
|q(x_{1})q(x_{2})q(x_{3})\rangle $. Using (\ref{p}) and assuming the flavor $%
SU(3)_{F}$ symmetry, it can be shown that
\begin{equation}
\begin{array}{c}
\langle r^{2}\rangle _{ch}=\langle p\uparrow
|\sum_{i=1}^{3}e_{i}r_{i}^{2}|p\uparrow \rangle \\
=\int d\Omega _{1}dr_{1}r_{1}^{4}\Psi _{u}^{\uparrow \dag }(x_{1})\Psi
_{u}^{\uparrow }(x_{1}) \\
=L^{2}\frac{\int_{0}^{\infty }dzz^{2}[G^{2}+H^{2}]}{\int_{0}^{\infty
}dz[G^{2}+H^{2}]}%
\end{array}
\label{charge}
\end{equation}%
where $\Psi _{u}^{\uparrow }(x_{1})$ is the space part of the wavefunction (%
\ref{psiQ}) of the quark(taken to be up quark here, for instance) at $x_{1}$%
, $e_{i}(i=1,2,3)$ stands for the quark charge($e_{u}=+2/3$, $e_{d}=-1/3$)
in the unit of electric charge $Q_{e}$. Numerically, the result is ($%
L=0.7574fm$)
\begin{equation*}
\langle r^{2}\rangle _{ch}=1.326L^{2}\simeq (0.872fm)^{2}.
\end{equation*}

Similarly, the magnetic moment can be shown to be (using the Dirac matrices $%
\mathbf{\hat{\alpha}}=\gamma ^{0}\mathbf{\gamma }$)
\begin{equation}
\begin{array}{c}
\mathbf{\mu }_{p}=\langle p\uparrow |\sum_{i=1}^{3}\frac{e_{i}}{2}(\mathbf{r}%
\times \mathbf{\hat{\alpha}})_{i}|p\uparrow \rangle  \\
=3\langle p\uparrow |\frac{e_{1}}{2}(\mathbf{r}\times \mathbf{\hat{\alpha}}%
)_{1}|p\uparrow \rangle  \\
=\frac{Q_{e}}{2}\int d\Omega _{1}dr_{1}r_{1}^{2}\psi _{u}^{\uparrow \dag
}(x_{1})\left(
\begin{array}{cc}
0 & \hat{r}\times \vec{\sigma} \\
\hat{r}\times \vec{\sigma} & 0%
\end{array}%
\right) \psi _{u}^{\uparrow }(x_{1}),%
\end{array}
\label{mag}
\end{equation}%
which, upon using (\ref{psiQ}), becomes,
\begin{equation}
\begin{array}{c}
\mathbf{\mu }_{p}=\frac{Q_{e}N^{2}}{2}\hat{z}\int_{0}^{\infty
}drrG(r)H(r)\int_{-1}^{1}\sin ^{2}\theta d\cos \theta , \\
=\frac{2Q_{e}N^{2}L^{2}}{3}\hat{z}\int_{0}^{\infty }dzzG(z)H(z), \\
=\frac{2Q_{e}L}{3}\hat{z}\frac{\int_{0}^{\infty }dzzG(z)H(z)}{%
\int_{0}^{\infty }dz[G^{2}+H^{2}]}.%
\end{array}
\label{mGF}
\end{equation}%
The numerical result for $\mu _{p}$(in the unit of the nuclear magneton $\mu
_{N}=Q_{e}/2m_{p}$) yields
\begin{equation*}
\frac{\mu _{p}}{\mu _{N}}=\frac{4m_{p}L}{3}\frac{\int_{0}^{\infty
}dzzG(z)H(z)}{\int_{0}^{\infty }dz[G^{2}+H^{2}]}\simeq 2.704.
\end{equation*}

In \textrm{Table I }we present our report for the calculations, being
compared with the experimental data(Exp.) as well as the calculations by
other models, including the MIT bag model, the Skyrme model\cite{ANW}, the
HCB model\cite{JAC87}, and the (perturbative) chiral quark model(PCQM)\cite%
{PCQM}. A nice agreement with the data is obtained.

We note that though (\ref{CQ1}) is motivated by the chiral bag model, it is
in the range of the chiral quark models\cite{PCQM} where the confinement is
put in phenomenologically, and chiral symmetry is implemented by
construction. The running of the valence quark mass agrees with asymptotic
freedom of QCD in principle. The our model (\ref{CQ1}) imitates the Skyrme
model when the confining scale(size of the quark core) $a\rightarrow 0$,
while it tends to the chiral bag model when $S(x)$ becomes a vallum-like
potential.

\begin{tabular}[t]{ccccccc}
\multicolumn{7}{c}{Table I} \\ \hline
Quantities & MIT\cite{DeGrand75} & Skyrme\cite{ANW} & HCB\cite{JAC87} & PCQM%
\cite{PCQM} & This Work & Exp. \\ \hline
$R[fm]$ & $1.0$ & $1.0$ & $0.6$ & $0.55\sim 0.65$ & $0.909$ &  \\
$e$ & [$\alpha c=2.2$] & $5.45$ & $4.5$ &  & $2.80$ &  \\
$f_{\pi }[fm]$ & [$Z=1.84$] & $93$ & $93$ & $88$ & $93$ & $93$ \\
$a[GeV^{-1}]$ &  &  &  &  & $0.296$ &  \\
$B^{1/4}[MeV]$ & $146$ &  & $150$ & [$B_{0}=1400$] & $134$ &  \\
$m_{p}[MeV]$ & $938$ & $939$ & $1425$ & $938.3$ & $938.27$ & $938.27$ \\
$\langle r^{2}\rangle _{ch}^{1/2}[fm]$ & $0.73$ & $0.59$ & $0.48$ & $0.85$ &
$0.8723$ & $0.877$ \\
$|\mu _{p}|[\mu _{N}]$ & $1.93$ & $1.87$ & $2.19$ & $2.6$ & $2.704$ & $2.793$
\\ \hline
\end{tabular}

The main features of model (\ref{CQ1}), which differs from the chiral quark
model, and the bag models as well as the Skyrme models mentioned above, lies
in

(i) It identifies the dynamical role of the term $\sim tr[2\pm (U+U^{\dag
})] $ as a bag-like function, which breaks explicitly chiral symmetry of the
model, as it does in ChPT\cite{ChPT}.

(ii) The generation of the valence quark mass via the pion dressing is made
explicit, in a way that is consistent with the chiral symmetry and breaking
of QCD, and that the valence quark mass vanishes at short distance.

(iii) The bag radius($\sim 0.9fm$) is much bigger than the quark-core size ($%
\sim 0.3fm$), which emphasizes the role of pion, but agrees with the two
scales of QCD: $\Lambda _{QCD}^{-1}=0.6\sim 1fm$ and $\Lambda _{\chi
}^{-1}\sim 0.2fm$.

\section{Summary}

Motivated by the hybrid chiral bag model, we propose a bag-like function of
nonlinear pion to make the bag surface of baryon dynamical. Assuming quark
confinement, we illustrated that the softening of the bag surface via
nonlinear pion gives rise to a chiral quark model with an
effectively-generated mass of quark, in which the confined quark moves in a
background of nonlinear pion in a chiral invariant way. The running
effective mass of quark is obtained through renormalization of the quark
field in terms of pion, and a relation for bag constant $B$ $\simeq 2f_{\pi
}^{2}m_{\pi }^{2}$ is obtained$\allowbreak $. Some static properties(mass,
charge radius, magnetic moment) of proton are computed by solving model with
a trial Skyrmion profile, in good agreement with the experimental data. The
bag radius are estimated to be about $0.91fm$ and the size of quark core to
be about $0.3fm$.

\section{Acknowledgements}

D. J thanks YuBin Dong, Jun He, Xiang Liu, and Qiang Zhao for discussions.
This work is supported by the National Natural Science Foundation of China
(No.11265014) and (No.10965005).

\end{document}